\documentclass{aa}  
\usepackage{graphicx}
\begin{document}
\title{The Galaxy Mass Function up to z=4 in the GOODS-MUSIC
   sample: into the epoch of formation of massive galaxies
\thanks{The observed mass functions are
available in electronic form
at {\sf http://lbc.oa-roma.inaf.it/goods/massfunction}}
}

   \author{A. Fontana\inst{1}
          \and
          S. Salimbeni\inst{1}
          \and
	  A. Grazian \inst{1}
          \and
          E. Giallongo\inst{1}
          \and
          L. Pentericci \inst{1}
          \and
          M. Nonino \inst{2}
          \and
	  F. Fontanot \inst{3}
	  \and
          N. Menci \inst{1}
          \and
          P. Monaco \inst{3,2}
          \and
          S. Cristiani\inst{2}
          \and
          E. Vanzella\inst{2}
          \and
          C. De Santis \inst{1}
          \and
          S. Gallozzi \inst{1}
          }

   \offprints{A. Fontana, \email{fontana@mporzio.astro.it}}

\institute{INAF - Osservatorio Astronomico di Roma, Via Frascati 33,
I--00040 Monteporzio (RM), Italy
\and INAF - Osservatorio Astronomico di Trieste, Via G.B. Tiepolo 11,
I--34131 Trieste, Italy
\and Dipartimento di Astronomia, Universit\`a di Trieste, Via G.B. Tiepolo 11,
I--34131 Trieste, Italy}

   \date{Received 21 April 2006 ; accepted 21 August 2006}
   \titlerunning{GOODS-MUSIC: the Galaxy Mass Function}
 
  \abstract
{}
{  
The goal of this work is to measure the evolution
of the Galaxy Stellar Mass Function and of the resulting Stellar Mass
Density up to redshift $\simeq 4$, in order to study the assembly of 
massive galaxies in the high redshift Universe.  } 
{We have used the GOODS-MUSIC catalog, containing $\sim$3000
Ks-selected galaxies with multi-wavelength coverage extending from the
U band to the Spitzer $8 \mu$m band, of which 27\% have spectroscopic
redshifts and the remaining fraction have accurate photometric
redshifts. On this sample we have applied a
standard fitting procedure to measure stellar masses. We 
compute the Galaxy Stellar Mass Function and the resulting Stellar
Mass Density up to redshift $\simeq 4$, taking into proper account the
biases and incompleteness effects. }
{Within the well known trend of global decline of the Stellar Mass
Density with redshift, we show that the decline of the more massive
galaxies may be described by an exponential timescale of $\simeq
6$~Gyrs up to $z\simeq 1.5$, and proceeds much faster thereafter,
with an exponential timescale of $\simeq 0.6$~Gyrs.  We also show that
there is some evidence for a differential evolution of the Galaxy
Stellar Mass Function, with low mass galaxies evolving faster than
more massive ones up to $z\simeq 1-1.5$ and that the Galaxy Stellar
Mass Function remains remarkably flat (i.e. with a slope close to the
local one) up to $z\simeq 1-1.3$. }
{The observed behaviour of the Galaxy Stellar Mass Function is
consistent with a scenario where about 50\% of present--day massive
galaxies formed at a vigorous rate in the epoch between
redshift 4 and 1.5, followed by a milder evolution until the
present-day epoch.}
\keywords{Galaxies:distances and redshift - Galaxies: evolution -
Galaxies: high redshift - Galaxies: fundamental parameters -
Galaxies: mass function}
\maketitle
%

\section{Introduction}

The observational evidence of the continuous increase of the
stellar content of galaxies over cosmic times has emerged only recently from
a series of observations and surveys which have made use of new
sensitive IR instrumentation.  Following  the first pioneering
studies (\cite{giallongo98}, \cite{brinchmann00}, \cite{papovich01})
that set-up the technique for estimating stellar masses in high
redshift galaxies, several surveys pointed out that the global stellar
content of the Universe, as measured by the average stellar mass
density, increases with cosmic time (\cite{brinchmann00},
\cite{dickinson03}, \cite{fontana03}, \cite{rudnick03},
\cite{glazebrook04}, \cite{drory04}, \cite{fontana04}, F04 hereafter, 
\cite{rudnick06}).

While the average stellar mass density provides a global picture of
the process of stellar assembly, the Galaxy Stellar Mass Function
(GSMF in the following) provides a more detailed view on how this
process evolves as a function of the galaxy mass itself. At low
redshift, accurate GSMF have been obtained from the 2dF
(\cite{Cole01}) or 2MASS--SDSS (\cite{bell03}) surveys.  Because of
the large area and of the accuracy in the redshift estimate that is
needed to compile a GSMF, very few GSMFs have been obtained so far.
The MUNICS survey (\cite{drory04}) and the K20 survey (F04) first
explored the evolution of the GSMF up to $z\simeq 1-2$ in a somewhat
complementary fashion. The former adopted a wide, relatively deep
sample on a wide area ($\simeq 5000$ objects at $K\leq19$),
mostly relying on photometric redshifts, sampling the GSMF up to
$z\simeq 1.5$, while the latter adopted a deeper but smaller sample
($\simeq 500$ objects at $K\leq20$), with excellent spectroscopic
coverage, extending up to $z=2$. In the overlapping redshift ranges,
the two GSMF agree quite well, and both analyses suggested a decline
in the density of massive galaxies at $z\simeq 1$, of the order of
50-70\% with respect to the local value. Recently, \cite{drory05}
extended such analysis to higher ($z\simeq 5$) redshift, adopting a
combination of $I$ and $K$--selected samples.

F04 also pointed out a tentative evidence of a differential
evolution of the GSMF, such that massive galaxies appear to evolve
less than low mass galaxies, at least up to $z\simeq 1.5$. As we shall
show in the following, this trend is also shown by our new data.

Along a different line, evidence has emerged that the GSMF for
galaxies of different spectral or morphological types shows a
significant evolution, with an increase in the fraction of stellar
mass residing in late type galaxies at $z\simeq 1-1.5$ with respect to low
redshift (F04, \cite{bundy05}, \cite{franceschini06}).

Although these surveys have already provided a first picture of the
evolution of the GSMF, the uncertainties involved in this exercise are
still large.  In particular, the exploration of the high redshift
Universe is still very limited. In several cases, the lack of
long--wavelength data hampered the estimate of the stellar mass at
high redshift, for which an adequate sampling of the rest frame
optical--near infrared part of the spectrum is essential. In addition,
the collection of a large sample of high redshift galaxies requires a
combination of large areas and deep near--IR observations. For
different reasons, all the surveys mentioned above suffer of
these limitations, although to a different extent.

In this context, the GOODS-South survey provides an excellent
opportunity of improving in a significant way our knowledge of the
high--$z$ GSMF. The combination of deep, wide IR observations with
Spitzer (in its four channels from 3.5 to 8~$\mu$m) and VLT-ISAAC (in
$J$, $H$ and $Ks$), coupled with high quality imaging data in the
optical domain (both ACS and VLT--VIMOS), and of a large, extended
spectroscopic coverage make the GOODS-South field ideal for this
investigation. From this
public data set we have obtained a multicolour catalog of faint
galaxies, that we named GOODS--MUSIC (GOODS MUltiwavelength Southern Infrared
Catalog), that we describe in \cite{grazian06a}.

In this work we use the GOODS--MUSIC sample to improve the previous
estimates of the GSMF in several ways. The most important is that our analysis
includes the $3.5-8 \mu$m Spitzer observations of the complete
$Ks$--selected data set. In addition, it extends to $Ks$ magnitudes
deeper than any previous survey, enabling us to obtain a complete
sample of galaxies up to $z\simeq 4$, and to measure the slope of the
low mass side of the Galaxy Stellar Mass Function up to $z\simeq 1.3$.
Compared to our previous analysis of the K20 survey data, the present
data set provides a final sample that is 6$\times$ larger than the K20
one, on which we adopt a more sophisticated technique to parametrize
the evolution of the Galaxy Stellar Mass Function with a
redshift-dependent Schechter fit, that provides interesting clues on
the evolution of massive galaxies.

The paper is organized as follows. In Sect.2, after reminding the
basic feature of our dataset and of the procedure that we adopt to
extract the stellar masses from each galaxy, we discuss how the
inclusion of the Spitzer bands affects the mass estimates and
highlight the selection effects that will affect our analysis. In
Sect.3, we present the basic results of our analysis, namely the
Galaxy Stellar Mass Function and the resulting mass density, both in a
binned as well as in a parametric fashion. In Sect.4, we compare our
basic findings with the prediction of recent theoretical
models. Finally, in Sect.5, we focus on the highest redshift range, to
discuss the reliability of the photometric redshifts on a few,
intriguing objects that might be at very high redshift. The discussion and
conclusions are summarized in Sect.6.

All the magnitudes cited below are in the AB system. To scale
luminosities and compute volumes we have adopted a ``concordance''
cosmological scenario with $\Omega_\Lambda = 0.7$ and
$H_0=70$~km~s$^{-1}$~Mpc$^{-1}$.

\section{Stellar Masses in the GOODS--MUSIC sample}
\subsection{The Data}
We use GOODS-MUSIC, a multicolour catalog extracted from the deep and
wide survey
conducted over the Chandra Deep Field South (CDFS) in the framework of
the GOODS public survey.  The procedures that we adopted are described
at length in \cite{grazian06a}. We remind here only the basic
features.

The data comprise a combination of images that extend over 14 bands,
namely U--band data from the 2.2ESO ($U_{35}$ and $U_{38}$) and
VLT-VIMOS ($U_{VIMOS}$), the $F435W$, $F606W$, $F775W$ and $F850LP$
($Z_{850}$) ACS images, the $JHKs$ VLT data and the Spitzer data provided by
IRAC instrument (3.6, 4.5, 5.8 and 8.0 $\mu$m). From this dataset, we
have obtained a multiwavelength catalog of 14847 objects, selected
either in the $Z_{850}$ and/or in the $Ks$ band.  For the purposes of the
present work, we will mainly use the $Ks$--selected catalog. This
consists of 2931 galaxies (after removal of known or candidate AGNs
and Galactic stars), 1922 of which have $U_{VIMOS}$ coverage, 1762
have $H$ coverage, and all have a complete coverage in the remaining
12 bands, most notably including the Spitzer ones. Since the detection
mosaics have an inhomogeneous depth, we have divided the $Ks$ sample
into 6 independent catalogs, each with a well defined magnitude limit
and area, that we use to compute mass functions and mass densities.
However, the largest fraction of the sample has a typical magnitude
limit of $Ks\simeq 23.5$ that we will adopt for more qualitative
arguments.

Colours have been measured using a specific software for the accurate
``PSF--matching'' of space and ground based images of different
resolution and depth, that we have named ConvPhot (De Santis et al
2006).  We have cross correlated our catalog with the whole
spectroscopic catalogs available to date, from a list of surveys,
assigning a spectroscopic redshift to more than 1000 sources.  We note
that in this work we use a spectroscopic sample that is wider than
that presented in \cite{grazian06a}, thanks to the increased number
of spectra publicly available (\cite{vanzella05}).  Finally, we have
applied our photometric redshift code, developed and tested over the
years in a series of works (\cite{fontana00}, \cite{cimatti02},
\cite{fontana03}, \cite{fontana04}, \cite{giallongo05}), that adopts a
standard $\chi^2$ minimization over a large set of templates obtained
from synthetic spectral models. The comparison with the spectroscopic
sample (\cite{grazian06a}, \cite{grazian06b}) shows that the
quality of the resulting photometric redshifts is excellent, with a
r.m.s. scatter in $\Delta z/(1+z)$ of 0.03 and no systematic offset.

In summary, the final sample adopted in this work consists of 2931
galaxies, complete to a typical magnitude of $Ks\simeq 23.5$, over an
area of 143.2 sq. arcmin, 815 of which with reliable spectroscopic
redshift (i.e. 28\% of the total sample) and the remaining fraction
with well tested 14 bands photometric redshifts.

\subsection{Stellar Masses in the Spitzer era}
\begin{figure}
\includegraphics[width=9cm]{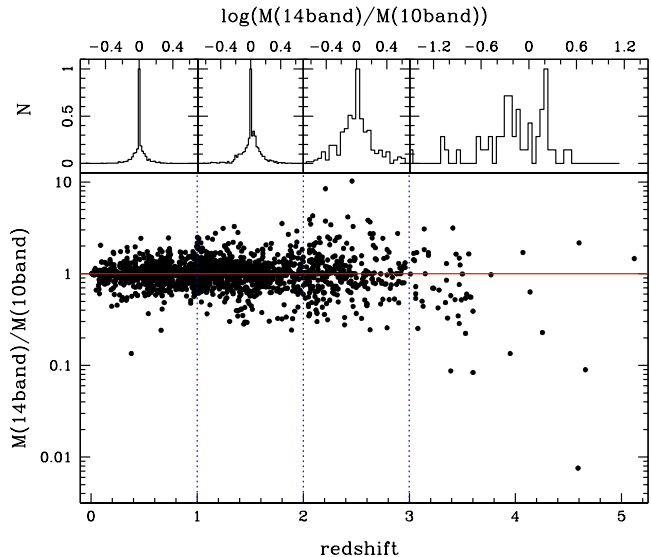}
\caption{Relation between the stellar masses derived with the
inclusion of the Spitzer bands (14 bands) and those without the
Spitzer bands (10 bands), as a function of redshift, on the
Ks--selected galaxies of the GOODS-MUSIC sample.  In the lower panel we
plot the ratio between the two estimates, for each galaxy in the
$Ks$--selected sample. In the upper, the distribution of such a ratio
in different redshift ranges.}
\label{spitzer}
\end{figure}

The method that we applied to estimate the galaxy stellar masses on
this dataset is exactly the same that we developed in previous papers
(\cite{fontana03}, F04), and similar to those adopted by other groups
in the literature (e.g. \cite{dickinson03}, \cite{drory04}). 
Briefly, it is based on a set of templates, computed with standard
spectral synthesis models (Bruzual \& Charlot 2003 in our case),
chosen to broadly encompass the variety of star--formation histories,
metallicities and extinction of real galaxies.  To compare with
previous works, we have used the Salpeter IMF, ranging over a set of
metallicities (from $Z=0.02 Z_\odot$ to $Z=2.5 Z_\odot$) and dust
extinction ($0<E(B-V)<1.1$, with a Calzetti extinction curve). Details
are given in Table 1 of F04. For each model of this grid, we have
computed the expected magnitudes in our filter set, and found the
best--fitting template with a standard $\chi^2$ normalization. The
stellar mass and other best--fit parameters of the galaxy are found
after scaling to the actual luminosity of the observed galaxy. Pros
and cons of the method have been discussed in several papers
(\cite{papovich01}, \cite{fontana04}, \cite{shapley04}), and we refer
to them for a detailed discussion of the systematics involved in this
exercise.

\begin{figure}
\includegraphics[width=9cm]{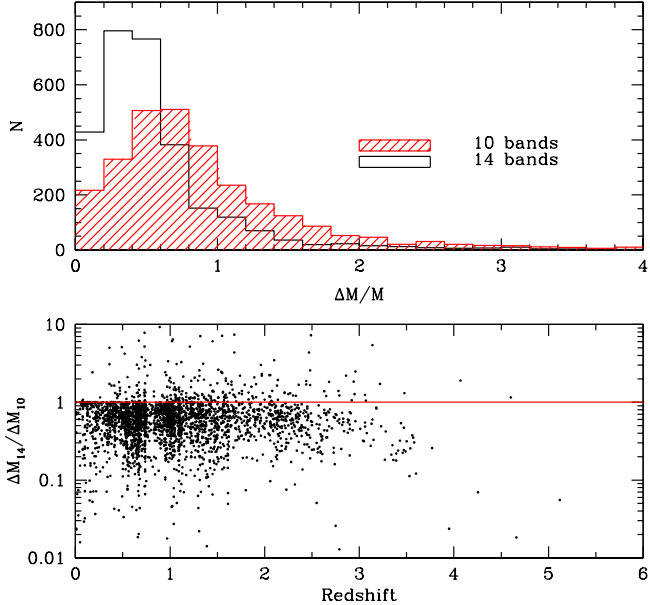}
\caption{Effect of the Spitzer bands on the mass estimate.  {\it Upper
panel} we plot the distribution of the relative uncertainty (at the
$1\sigma$ level) in the mass estimate, when the Spitzer bands are
included (14 bands) or not (10 bands). The distribution is computed
over all $Ks$--selected galaxies. {\it Lower panel}: Ratio of the
uncertainty (at the $1\sigma$ level) in the mass estimate, when the
Spitzer bands are included (14 bands) or not (10 bands), as a function
of redshift.}
\label{mass_error}
\end{figure}

The major difference with all previous estimates of the GSMF arises
from the inclusion of the 4 Spitzer bands, longward of 2.2 $\mu$m. For
galaxies at $z>2$, these bands are essential to sample the spectral
distribution in the rest-frame optical and near-IR bands, that are
necessary to provide reliable constraints on the stellar mass. We
quantitatively assess their importance in Fig.\ref{spitzer}, where we
plot the ratio between the mass estimates with (14 bands) and without
(10 bands) the Spitzer bands on the sample of $Ks$--selected galaxies,
as function of redshift.

It is immediately appreciated that the inclusion of the Spitzer bands
does not provide statistically significant changes at $z<2$, where
most of the galaxies are fitted with the same best--fitting models in
both cases. The scatter in the mass estimate is entirely consistent
with the expected uncertainty due to model degeneracy.  At higher
redshift, the scatter increases significantly. At $z\simeq 2.5$, the
r.m.s. fluctuation is of about $0.23$, although with no systematic
shift. At higher redshift, the r.m.s. is as large as $0.45$ {\it dex},
with evidence of systematic overestimate (of 0.2 {\it dex}) of the
stellar mass when the Spitzer bands are not included. 

The same improvement can be found by looking at the uncertainty
in the mass estimate. As in F04, we computed the $1\sigma$ confidence
level on each mass estimate by scanning the $\chi^2$ levels, allowing
the redshift to change in case of objects with photometric
redshift. In Fig.\ref{mass_error} we compare these estimates when the
Spitzer bands are included (14 bands) or not (10 bands). As clearly
shown, the formal intrinsic uncertainty decreases from a typical value of
60\% to 40\% on the global sample, and decreases by a factor of three
for objects at $z>2$.

These results
confirm that, in the absence of the Spitzer bands, the
estimates of the stellar mass for galaxies at $z>3$ are very uncertain
and possibly biased, such that detailed astrophysical analysis based
on such estimates are likely premature.

\subsection{Stellar masses of the GOODS-MUSIC data sample}

\begin{figure}
\includegraphics[width=9cm]{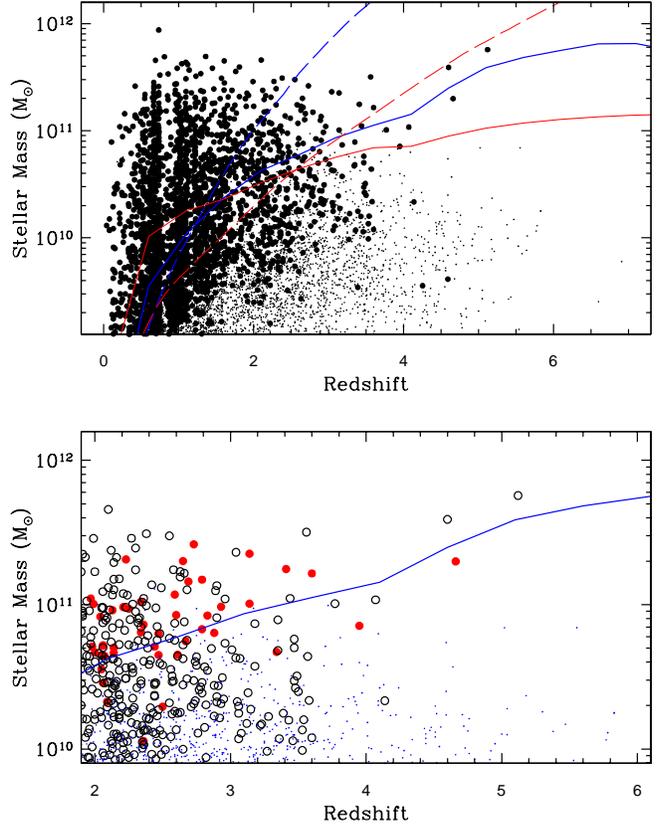}
\caption{Stellar Masses for individual galaxies of the GOODS-MUSIC
sample, as a function of redshift. {\it Upper panel}: the $Ks$--selected
(filled circles) and the $Z_{850}$--selected sample (small dots) on the
overall redshift range. {\it Lower panel}: same quantities, for galaxies at
$z\geq 2$.  Objects in the $Ks$--selected sample are shown with large
empty circles if they are also included in the $Z_{850}$--selected one, and
with large filled circles if included {\it only} in the $Ks$-band
catalog. Small dots are from the remaining $Z_{850}$--selected objects. In
both panels, lines represent the completeness threshold, computed as
described in the text. Thick lines correspond to the completeness
threshold for a $Ks$--selected sample with respect to passively
evolving models (continuous line) and dusty starburst with
$E(B-V)=1.1$ (dashed). The thin lines show the same thresholds if a
4.5$\mu$m--selected sample would be used, for the two same spectral
models.}
\label{massz}
\end{figure}

In Fig.\ref{massz} we plot the resulting stellar masses of the
GOODS-MUSIC sample, as a function of redshift. We include in the plot
both the objects of the $Ks$--selected sample as well as those of the
$Z_{850}$--selected one that have detected flux in the $Ks$ band.

One can immediately appreciate that massive galaxies (i.e. those above
the $z=0$ characteristic mass $M_*\simeq 10^{11.}M_\odot$) are
thoroughly detected up to at least $z\simeq 4$, and possibly even at
higher $z$, although the reliability of photometric redshifts for the
massive objects at $z\simeq5$ is uncertain, as discussed in Section 5.
We also note that such plot appears to be qualitatively different from
the analogous plot by \cite{drory05}, that contains several quite
massive galaxies (close to $10^{12}M_\odot$) up to $z\simeq 5$.  This
difference is likely due to a combination of effects, induced by the
different techniques applied for photometry and photometric redshifts,
the availability of an extended spectroscopic coverage, and probably
even more important the lack of Spitzer observations in the
\cite{drory05} sample. 

In the lower panel of Fig.\ref{massz} we focus on the high redshift
range of our sample, and we use different symbols to differentiate
among galaxies that are selected {\it only} in the $Ks$--selected
sample, those that are selected {\it both} in $Ks$ and $Z_{850}$ and
those that are selected in $Z_{850}$ only. As expected, galaxies that
are selected in $Z_{850}$ only, and hence have a very low flux in
$Ks$, populate the low mass region of the distribution. On the
contrary, a number of galaxies selected {\it only in $Ks$} contribute
to the population of massive galaxies at $z>2$: these are typically
very faint or sometimes even undetected in the $Z_{850}$--band,
resulting from highly dust--reddened or passively fading stellar
population at high redshift.  These objects significantly contribute
to the number density of massive systems. In a similar way, one
expects that other massive galaxies at these redshifts might be
included by selecting the sample at even higher
wavelengths, as those obtained with the Spitzer observations used
here.

\subsection{Incompleteness and selection criteria}
It is important to remark that our sample, as well as any other
magnitude -limited sample, does not have a well defined, sharp limit
in stellar mass. This incompleteness effect arise from the fact that
galaxies have a range of $M/L$ ratios, and can be illustrated as
follows. In the GOODS--MUSIC sample, at $z=1$ we find that galaxies
have a range of $M/L$ extending from 0.9 (for redder objects) to 0.046
(for bluer objects).  Assuming for simplicity a sharp magnitude limit
$Ks=23.5$, this corresponds to stellar masses from $M_\odot \simeq 1.5
\cdot 10^{10}$ (for redder objects) to $M_\odot \simeq 6 \cdot 10^8$
(for bluer objects). However, since galaxies have a range of $M/L$
ratios, additional objects with stellar masses in the range $6 \cdot
10^8 < M_\odot < 1.5 \cdot 10^{10}$ will lay at fainter fluxes,
$Ks>23.5$, which are not included in the present sample. Hence, the
census of galaxies in the range $6 \cdot 10^8 < M_\odot < 1.5 \cdot
10^{10}$ will definitely be incomplete in our sample, at $z\simeq
1$. A complete visualization of this effect can be obtained by looking
at Fig~16 of F04. Needless to say, this effect exists in any
magnitude-limited sample and at any redshift, although going to redder
wavelengths alleviates its impact.

This effect has been described at length by Dickinson et al 2003,
Fontana et al 2003 and F04, and can be dealt with in two different
ways. First, one can compute a limiting threshold in mass, such that
one can obtain a well defined mass--selected sample. All the relevant
statistics (mass densities, mass functions etc) should be obtained
using only the objects above such limit in stellar mass.

To better exploit the statistics of the sample, we have developed a
technique to correct for the incomplete coverage of $M/L$ ratios, that
allows to at least partially recover the fainter side of the
sample. This technique will be used to estimate the GSMF and will be
described in Sect. 3. In our sample, this technique allows to extend
the GSMF by about $0.4 dex$ in each redshift bin: the leading effect
remains therefore the conservative estimate of the completeness limit,
that we have derived as follows.

The limiting threshold in mass can be obtained by computing the
maximal $M/L$ that is allowable at each redshift and multiplying it by
the magnitude limit of the survey. Below such a threshold in mass, one
can still detect galaxies and measure their mass, but will start to
miss galaxies of the same mass but with lower luminosity (i.e. higher
$M/L$). Unfortunately, such a limiting threshold depends on the
assumed spectral properties for the targeted galaxies (and obviously
on the bandpass adopted to select the catalog).

In the simplest case, one can compute the maximal $M/L$ for passively
evolving systems.  In this case, it is easy to draw a selection
curve, that we plot in Fig.\ref{massz} (thick solid line), that we
computed with a maximally old, single burst model normalized to
$Ks=23.5$. However, large values of $M/L$ can also be found in heavily
extincted star--forming galaxies. To estimate this effect, we have
adopted a simple star--forming model with a variable amount of
extinction, adopting a Calzetti extinction law. It turns out that, for
$E(B-V)\simeq 0.5$, the corresponding selection curve roughly
corresponds to the ``passively evolving'' one, and we do not plot it
for simplicity. Increasing the amount of dust, the selection curve
shifts to higher masses. In Fig.\ref{massz} we plot the case for
$E(B-V)\simeq 1.1$ (thick dashed line), that is among the highest
observed in spectroscopically confirmed EROs (\cite{cimatti02}).

The result must be read as follows: our $Ks$--selected sample is
expected to be complete against passively evolving objects with mass
above the thick solid line in Fig.\ref{massz}, i.e. grossly with mass
$M_*\geq10^{11}M_\odot$ up to $z\simeq 4$; however, at this mass level
it becomes progressively biased against the detection of extremely
dusty objects, probably already at $z\geq 2$.

It is interesting to predict what would be the advantages of adopting
a Spitzer--selected sample, scaled to the expected sensitivity of the
GOODS survey. We have computed the same selection curves (for
passively evolving and dusty objects) adopting a magnitude limit at
$4.5\mu$m of $m_{4.5}\leq 23.5$, and we have overplotted them in
Fig.\ref{massz} as thin solid and dashed lines. A first result is that
the adoption of a $4.5\mu$m--selected sample would not allow to extend
the sampling of passively evolving galaxies at lower masses than in
our catalog, but rather to extend the selection of objects at the same
level of $M_*\simeq10^{11}M_\odot$ well beyond $z\simeq 4$. Probably
the most important effect is that the adoption of a
$4.5\mu$m--selected sample would allow the detection of heavily
extincted, dusty objects well beyond $z\simeq 2$, probably up to
$z\simeq 3$ at the $M_*\simeq10^{11}M_\odot$ level. Objects detected
in the $4.5\mu$m band, but very faint or even undetected in $Ks$, do
exist in the GOODS area (see Yan et al 2005 for a few cases):
unfortunately, their inclusion in the present analysis would require a
detailed analysis (most notably of the reliability of their
photometric redshift), that would go beyond the purposes of this work
and that we defer to a dedicated paper.

\section{The Galaxy Stellar Mass Function}

\begin{figure*}
\centering \resizebox{\hsize}{!}{
\includegraphics[angle=0]{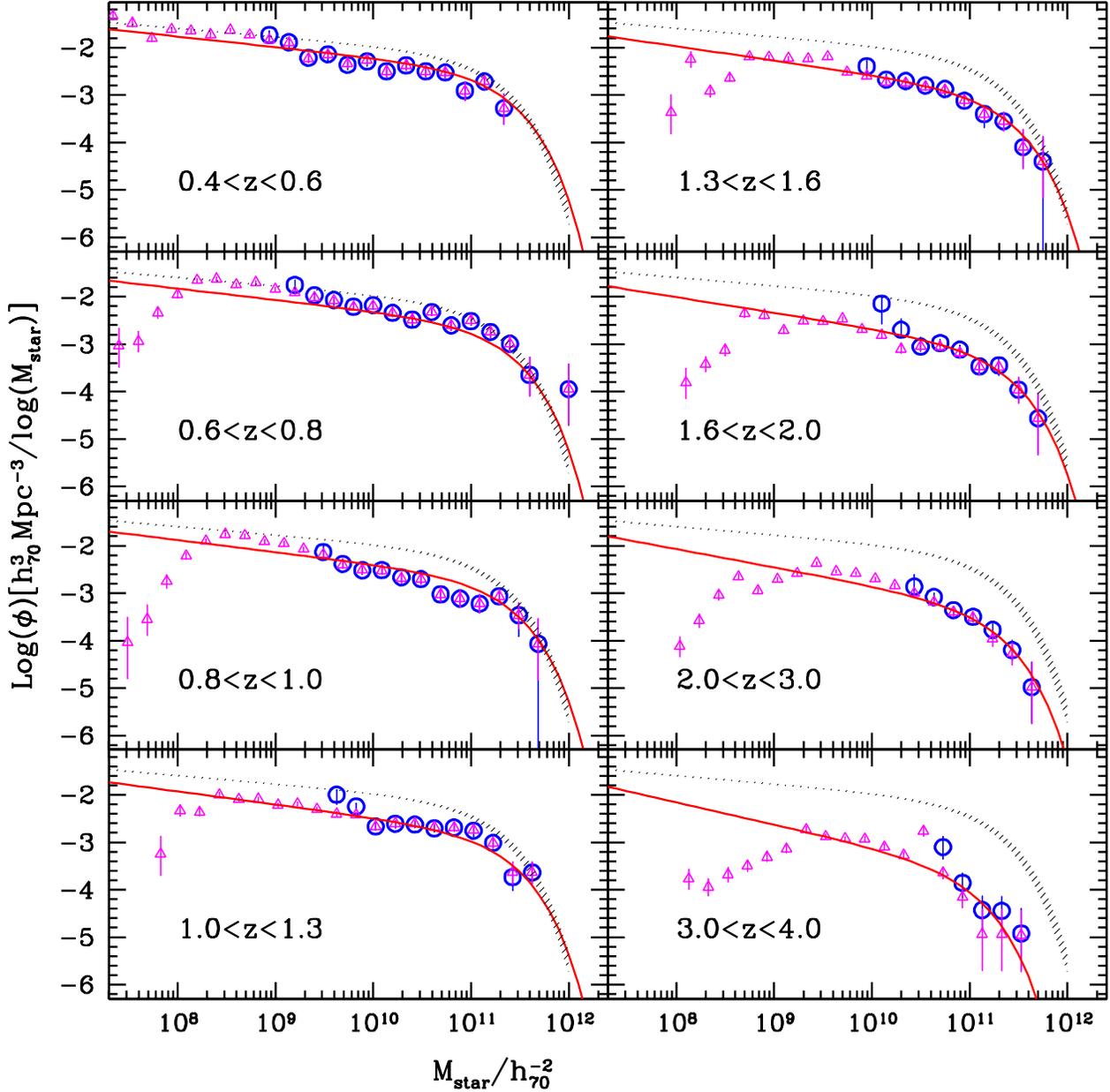}}
\caption{Galaxy stellar Mass Functions in the GOODS--MUSIC sample, in
different redshift ranges. Big circles represent the Galaxy Stellar
Mass Functions of the $Ks$--selected sample, computed with the
$1/V_{max}$ formalism up to the appropriate completeness level, as
described in the text, while small triangles show the Galaxy Stellar
Mass Functions of the $Z_{850}$--selected sample, computed without any
correction for incompleteness coverage of the $M/L$ at faint
limits. The dashed region represents the local GSMF of \cite{Cole01},
encompassing its original and the ``rescaled'' version that we obtained
in F04 (see text for details).
The solid line is the evolutionary STY fit of our data, computed over
the global redshift range $0.4<z<4$, that we describe in Section 3.2.
The observed mass functions are available in electronic form
at {\sf http://lbc.oa-roma.inaf.it/goods/massfunction}.}
\label{massfun}
\end{figure*}

\subsection{Computing the  Mass Function}
We have used the $Ks$--selected sample described above to compute
the GSMF at various redshifts. The GSMF is computed both by using the
standard $1/V_{max}$ formalism as well as by fitting a Schechter
function to the unbinned data. The results of the former method are
described in Sect. 3.2, while the latter in 3.3.

The GSMF must be computed taking into account the incompleteness
effects described above. To correct for this effect, we have
introduced a correction technique, which is carefully described in
F04, and that we remind briefly here. We start from the threshold
computed from passively evolving system, i.e. the thick solid line of
Fig.\ref{massz}: below such completeness threshold only a fraction
($f_{\rm obs}$) of objects of given mass will be observed. We then
obtain (at any redshift) the observed distribution of $M/L$ ratios for
objects close to the magnitude limit of the sample. Then, using such
distribution, we compute for each mass and redshift the fraction
$1-f_{\rm obs}$ of galaxies lost by effect of the incomplete coverage
of the $M_*/L_K$ ratio, by which we obtain $f_{\rm obs}$. As shown in
F04, Appendix B, relatively simple analytic formulae can be used to
describe this process. We adopted the same technique to the
GOODS-MUSIC sample, verifying that these simple analytic expression
still provide a good fit to the observed $M_*/L_K$ distribution,
despite the greater depth and area of the GOODS-MUSIC sample.  The
correction is applied only for a limited range of masses, where the
correction is less than 50\% (i.e. $f_{\rm obs}\geq 0.67$. This
correction factor is then applied to the volume element $V_{\rm max}$
of any galaxy in the $1/V_{\rm max}$ binned GSMF as well as in the
number of detected galaxies that enters in the Maximum Likelihood
analysis used to obtain the Schechter best-fits. As shown in Fig.~17
of F04, the inclusion of this treatment allows to extend the
computation of the GSMF to somewhat ($0.4 dex$) smaller masses than by
adopting the strict completeness limit. In contrast, neglecting the
whole treatment of incompleteness introduces a remarkable and
unphysical bending of the GSMF at low masses.

A further somewhat technical point is related to the choice of the
local GSMF that is used to estimate evolution.  As described in F04,
the parameter grid used as input to the spectral synthesis model
adopted here is different from the one adopted in the derivation of
the local GSMF by \cite{Cole01}, in particular for the absence of any
constraint on the galaxy ages. This systematic difference may lead to
a small but systematic bias in the measure of galaxy masses, that can
be described by an average relation
($\log(M)=1.027*\log(M_{cole})-0.3955$).  By applying this relation to
the original \cite{Cole01} local GSMF we obtain a ``rescaled'' version
of the local GSMF, where differences are in practice noticeable only
on the exponential tail of the GSMF. For sake of clarity, we will in
the following represent both the original \cite{Cole01} local GSMF as
well as its rescaled form that we derived in F04.

\subsection{The distribution of massive galaxies up to z$\simeq 4$}

The Galaxy Stellar Mass Function obtained in the GOODS--MUSIC sample
is shown in Fig.\ref{massfun}, where it is computed in several
redshift bins. We also plot in Fig.\ref{massfun} the GSMF obtained on
the $Z_{850}$--selected sample, naively computed with no correction
for the incomplete coverage of the $M/L$ distribution.  We explicitly
note that such a $Z_{850}$--selected GSMF is quite consistent with the
$Ks$--selected one at low and intermediate redshifts, before fading at
the lowest mass end because of the lack of correction for incompleteness.
Conversely, it misses a significant fraction of massive galaxies at
$z>2-3$, as expected in a $Z_{850}$--selected sample.

We also remark that the error bars of Fig.\ref{massfun} have been
computed with a full Monte Carlo simulation where we take into account
the redshift probability distribution of each galaxy in the sample.

Three different results can be inferred even from a visual
inspection of Fig.\ref{massfun}. First, the density of massive
galaxies clearly decreases with redshift, relatively mildly up to
$z\simeq 1.5$ and then much more convincingly at $z>2$. At $z\simeq
3.5$, their density is at least a factor 10$\times$ lower than in the
local Universe. This result confirm the trend that we already noted in
F04 (see also Drory et al 2005).  

 Second, the number density of lower mass system evolves
significantly with redshift (at $z\simeq 1$ the density of galaxies
with $M\simeq 10^{10}M_\odot$ is 4 times lower than the local one),
while the slope of the GSMF is relatively flat up to at least $z\simeq
1-1.5$, and does not appear to steepen significantly with respect to
the local one. The latter result is apparently at variance with the
results of in F04, where we found a significant steepening of the GSMF
at $z\simeq 0.7$, with the slope index $\alpha$ changing from
$\alpha=-1.18$ at $z=0$ to $\alpha=-1.4$ at $z=0.7$.  We first
carefully checked that no systematic effect is responsible for this
result: in particular, we verified that the GSMF computed with our new
sample in the very area of the K20 survey is comparable to our early
results (actually it is nearly perfectly coincident), despite the new
photometry of the whole sample.  We furthermore note that even in the
GOODS--MUSIC sample the slope of the GSMF is steeper than the average
in the redshift bin $z=0.8-1$, as also shown by the parametric
analysis presented in Sect.3.3, probably due to a cosmic variance
effect.  Taking also into account the agreement with the
$Z_{850}$--selected GSMF, we conclude that our new sample, that is
much deeper and wider than the K20 sample, allows to better represent
the global evolution of the slope of the GSMF with redshift.
 
Finally, there is some evidence of a differential evolution of the
GSMF, in the sense that more massive galaxies appear to evolve less
than low mass galaxies, at least up to $z\simeq 1.5$. This trend was
already noted in F04, but the more limited statistics prevented a
robust quantitative conclusion.

Although these conclusions are clearly supported by our data, we are
aware that the $1/V_{max}$ approach is very sensitive to the biases
induced by Large Scale Structures (LSS) and clustering, and that our
sample is clearly affected by these effects.  A well known effect is
the apparent steepening of the faintest points in case of clustering
(\cite{heyl97}): this is present in our sample at $1<z<1.3$ and
$1.6<z<2$, due to two large overdensities at the lower limit of the
bin. Such effects are amplified by the arbitrariness in the choice of
the redshift bins, and by our choice of keeping a small binning factor
in mass.  If one compares our sample with the local GSMF of
\cite{Cole01}, it is possible to see that fluctuations in the number
of massive galaxies exist, superimposed to the global evolution of the
GSMF.  In particular, the number of massive galaxies and even the
detection of galaxies with $M_*>10^{11}M_\odot$ is higher in the
redshift bins where LSS have been detected in the CDFS field, e.g. at
$z=0.67, 0.73$, $z\simeq 1.61$ (\cite{gilli03}, \cite{vanzella04},
\cite{vanzella05}), while our sample is relatively devoid of massive
galaxies at $z\simeq 0.9$, where an underdensity of galaxies exists in
the CDFS (\cite{perez05}). To overcome these limitations, we have
developed the STY analysis that we shall describe in the following
section.

\subsection{The STY approach}

An independent approach to the evaluation of the GSMF is the STY
fitting method (\cite{sandage}), mutuated from the corresponding
formalism of the Luminosity Function. This method assumes an analytic
expression of the GSMF and derives the free parameters with a Maximum
Likelihood analysis. It is less sensitive to clustering and LSS and
provides more quantitative hints on the evolution of the global
population. To minimize the impact of LSS, and to avoid any
arbitrariness in the definition of the redshift bins, we use here a
global, redshift-dependent parametrization of the GSMF. At each
redshift, the number density $\phi(M,z)$ of galaxies with mass
$M_*$ is assumed to have the functional form of a
Schechter function,

\begin{equation}
\phi^*(z) \log{(10)} [10^{(M-M^*(z))}]^{(1+\alpha^*(z))} \exp(-10^{(M-M^*(z)}))
\end{equation}

where $M=log_{10}(M_*/M_\odot)$ and the free parameters $\phi^*(z)$,
$\alpha^*(z)$ and $M^*(z)$ are functions of redshift.  For these
parameters, we have found that the following simple relations provide
an adequate fit to the overall evolution, in the redshift range
$0.4\leq z \leq 4$:

\begin{equation}
\phi^*(z)=\phi^*_0 \cdot (1+z)^{\phi^*_1}
\label{sche_phi}
\end{equation}

\begin{equation}
\alpha^*(z) = \alpha^*_0+ \alpha^*_1 \cdot z
\label{sche_alpha}
\end{equation}

\begin{equation}
M^*(z) = M^*_0+M^*_1 \cdot z+ M^*_2\cdot z^2 \\
\label{sche_mstar}
\end{equation}

where $\phi^*_0$, $\phi^*_1$, $\alpha^*_0$, $\alpha^*_1$, $M^*_0$, $M^*_1$,
$M^*_2$ are free parameters.

\begin{figure}
\centering \resizebox{\hsize}{!}{
\includegraphics[angle=0]{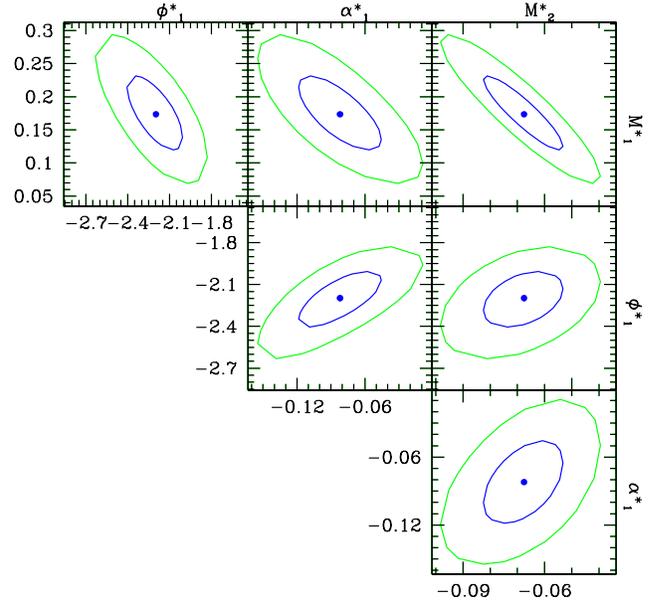}}
\caption{Contour levels of all the pairs of free parameters involved
in the Schechter fitting of the GSMF, as obtained by a Maximum
Likelihood minimization. Inner ellipses represent the $1\sigma$
contour, outer ones the $3\sigma$ contour. All contours have been
obtained with marginalization of the other free parameters.}
\label{contours}
\end{figure}

\begin{figure}
\centering \resizebox{\hsize}{!}{
\includegraphics[angle=0]{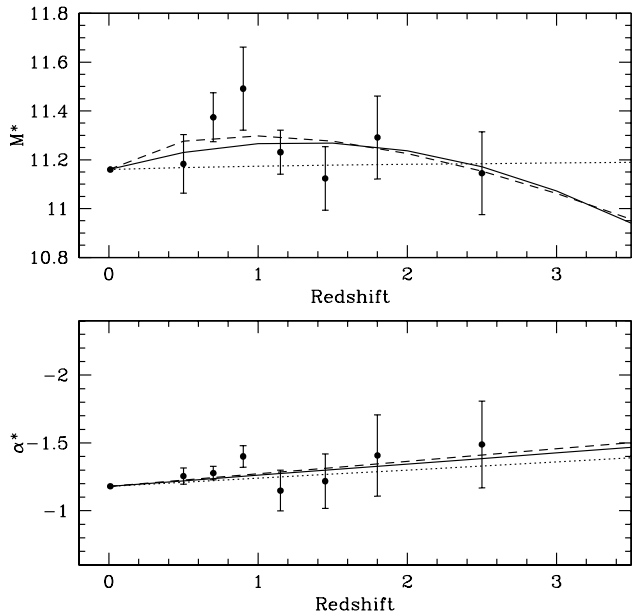}}
\caption{The evolution of the Schechter parameters $M^*$
(characteristic mass) and $\alpha$ (slope) as a function of
redshift. Points represent the individual fits to the single redshift
bin of Fig.\ref{massfun} with their uncertainty, while lines
represents the evolutionary STY fit of our data, computed over the
global redshift range $0.4<z<4$, that we describe in Section
3.2. Solid line is the second-order form for $M^*$ (our fiducial
model); dashed line is for $M^*=M^*_1 \times z + M^*_2 \times z^{M^*_3}$;
dotted line is for $M^*\propto \log(1+z)$. }
\label{sche_par}
\end{figure}

In this formalism, the three zero-th order parameters ($M_0^*$,
$\phi_0^*$ and $\alpha_0^*$) should ideally reproduce the local
values, as estimated for instance by \cite{Cole01}.  For this
reason we fixed the local characteristic mass and slope to the local
values of \cite{Cole01} and derived the other parameters. The resulting
fit is displayed in Fig.\ref{massfun} as a solid line, and the
relevant parameters are listed in Table \ref{tabSTY}. The Maximum
Likelihood analysis allows also to perform an estimate of the error
budget on the fitted parameters: we show in Fig.\ref{contours} the
$1\sigma$ and $3\sigma$ contour levels on each pair of free parameters,
obtained after marginalization of the other parameters. It is clearly
shown that all parameters are reasonably well constrained, in
particular those ($M^*_1, M^*_2$ and $\alpha^*_1$) that are more
important for the physical conclusions that we will draw in the
following. The resulting evolution with redshift of the characteristic
mass $M^*(z)$ and of the slope $\alpha^*(z)$ is shown in
Fig.\ref{sche_par}, where they are also compared with the
corresponding values obtained by a Schechter fit within each redshift
bin.

A first robust result of the STY analysis is that the slope of GSMF is
remarkably flat and with a small redshift evolution: up to $z\simeq
1-1.3$, the highest redshift where the slope is reliably measured in
our sample, the slope $\alpha$ changes of about $0.1$ in our fiducial
best--fit model. As discussed before, this is accompanied by a
sensible decrease in the number density of low mass galaxies, that for
galaxies with $M\simeq 10^{10}M_\odot$ is 4 times lower than the local
one.

The evolution of the characteristic mass $M^*(z)$ is more
complex. According to the best--fit values, the evolving $M^*(z)$ is
fitted with a second order law with negative curvature ($M^*_2 =
-0.07\pm0.02$) and a peak at about $z\simeq 1.35$, such that the
characteristic mass of the GSMF initially {\it increases}, reaches a
maximum around $z\simeq 1.35$ and than decreases. 

We carefully verified that this behaviour is not a spurious result of
our choice of the functional form of Eq.\ref{sche_mstar}, where the
second order law for $M^*(z)$ might introduce such a non-monotonic
behaviour.  First, as shown in Fig.\ref{contours}, the second order
term $M_2$ is definitely negative, suggesting that $M^*(z)$ actually
increases up to $z\simeq 1-1.3$ and decreases thereafter.  Such a
behaviour is also consistent with the trend observed in the values of
$M^*$ in the individual redshift bins, as shown in
Fig.\ref{sche_par}. More important, we arrived at this choice after
discarding simpler solutions.  In particular, we tested a linear form
for $M^*(z)$ (i.e. with $M_2^*=0$), as well as a more traditional
logarithmic function ($M^*(z) = M^*_0+M^*_1 \cdot \log(1+z)$). The
output of the latter form is also shown in Fig.\ref{sche_par}. In all
these cases, we were not able to fit the massive side of the GSMF at
intermediate redshifts ($0.6\leq z \leq 1.5$), that is systematically
underfitted, by nearly 50\%. As a further test, we have also allowed
for the exponent of the high order term to be free (i.e. $M^*(z) =
M^*_0+M^*_1 \cdot z+ M^*_2\cdot z^{ M^*_3}$), and found that it is
definitely larger than 1 ($M^*_3=1.2\pm0.05$), providing an evolution
that is similar to our fiducial model, as shown by the dashed line in
Fig.\ref{sche_par}. We conclude that such evolutionary form is a
robust property of our sample, and we shall adopt the second order
form for $M^*(z)$ shown in Eq. 4 as our best--fit fiducial model.

Other warnings arise from possible biases in our sample.  The most
obvious sources of biases are sample variance and the use of
photometric redshifts for a relatively large fraction of the
sample. We expect the latter uncertainty to be minor, given the large
spectroscopic coverage of the GOODS survey for the brightest galaxies
(where most of the evolution of $M_*$ is measured) and the good
accuracy that we achieve in photometric redshifts. Sample variance is
a more serious concern, since the GOODS--South field is definitely too
small to avoid spurious effects due to over- and under-densities along
the line of sight. In the redshift interval $0.6 < z < 1$ a range of
different densities exist, from overdensities (around $z=0.7$) to
underdensities (at $z\simeq 0.9$), and the increase of $M^*$ is
reassuringly seen in both. However, much wider surveys with comparable
depth are definitely required to confirm or dispute our finding.

\begin{figure}
\includegraphics[width=9cm]{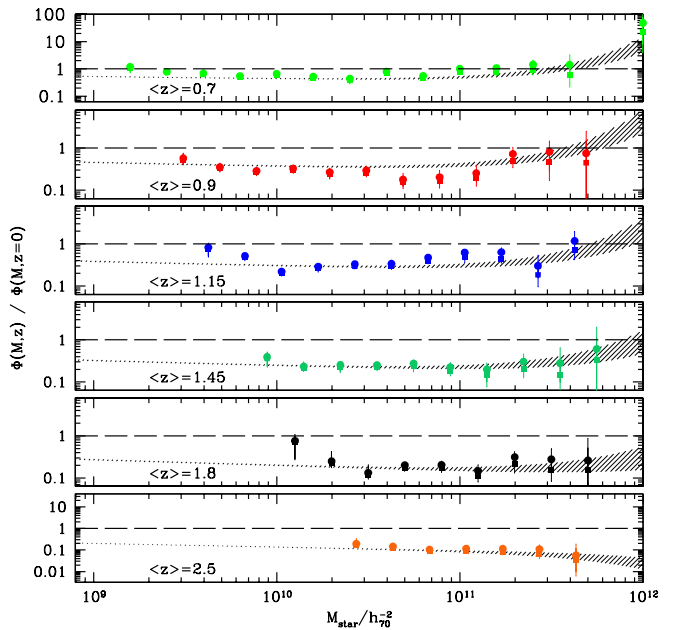}
\caption{Ratio between the observed Galaxy Stellar Mass Functions in 
the GOODS--MUSIC sample and the local GSMF, in the same redshift bins of
Fig.\ref{massfun}. The observed data are normalized to two different 
choices of the local
GSMF, namely the original one of \cite{Cole01} (filled squares)
and its rescaled version, that we  described in the text (filled circles).
The shaded area shows the corresponding ratios for the 
global Schechter fit, that we describe in Sect. 3.2.}
\label{mfdiff}
\end{figure}

Turning to physical interpretation of this result, we remark that the
evolution of the characteristic mass $M_*$ reflects an evolution of
the shape of the GSMF, that is progressively skewed toward larger
masses as it evolves from $z=0$ to $z\simeq 1.3$. The parametric
analysis of this section enforces what we already highlighted in
Sect.3.2, i.e. that the GSMF appear to have a differential evolution,
with more massive galaxies evolving less than low mass galaxies, up to
$z\simeq 1-1.3$. We note that this does not imply that the mass
density increases, since the increase of $M^*(z)$ is counterbalanced
by a decrease of the overall normalization $\phi^*(z)$.

The issue of the differential evolution of the GSMF is potentially
very important, since it is directly related to the ``downsizing''
scenario for galaxy evolution, such that it is worth a more careful
examination.  We plot in Fig.\ref{mfdiff} the ratio between our GSMF
and the local one, for different redshifts. In addition to the obvious
overall evolution of the GSMF with redshift, it is shown that the
density of galaxies above $10^{11}M_\odot$ - albeit decreasing -
remains closer to the local value up to $z\simeq 1.15$ than that of
the lower mass population. Such a trend appears to be progressively
stronger for the more massive galaxies of our survey, i.e. those with
a stellar mass in excess of $3\times 10^{11}M_\odot$. At larger
redshifts, however, this trend eventually breaks down, such that the
density of massive galaxies undergoes a strong evolution in the
redshift range $1.5<z<3$.

\subsection{The integrated Stellar Mass Density}
\begin{figure}
\includegraphics[width=9cm]{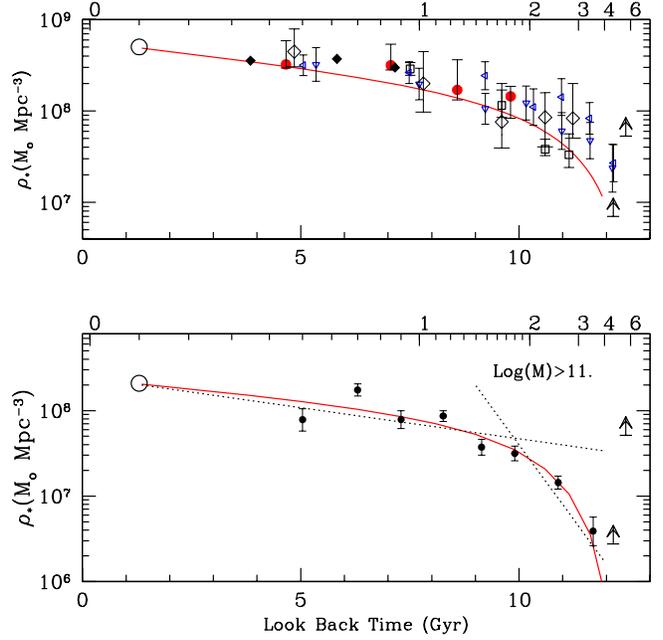}
\caption{Global Stellar  Mass Density  as function of  redshift.  {\it
Upper  panel}:  the  total   Stellar  Mass  Density  (integrated  from
log$(M_*/M_\odot)=8$  to log$(M_*/M_\odot)=13$)  as  estimated in  the
present  work, obtained by  an integration  of our  fiducial Schechter
function  (solid line). It  is compared  with other  estimates adopted
from the literature: the local value  by Cole et al. 2001 (large empty
circle), Fontana et al. 2004 (red filled circles), Fontana et al. 2003
(HDFS, open diamonds), Dickinson et  al. 2003 (open squares), Cohen et
al. 2001 (open crosses),  Brinchmann and Ellis 2000 (filled diamonds),
Drory et  al. 2005 (open triangles, K-selected  sample with horizontal
basis,  I-selected  sample with  vertical  basis).   To avoid  further
confusion,  we do  not plot  here the  mass density  computed  in this
paper. It  is useful to stress  that the solid  curve is not a  fit to
these  data, but  only our  best parameterization  of the  SMD  in the
GOODS-MUSIC sample, which agrees well with that of Fontana et al. 2004
in the K20 region and with Drory et al. 2005 on the GOODS-South field.
{\it Lower  panel:} the  mass density computed  only on  galaxies more
massive  than  log($M_*/M_\odot)=11$.  We  remark  that  the value  at
$z\simeq  0.5$ is  affected by an undersampling  of  massive galaxies
because of the small size of the volume element.  The solid line shows
the mass  density computed with  our fiducial Schechter fit.  See text
and Tab.\ref{tabSTY}  for details. The  two dotted lines  represent an
exponentially declining  evolution of  the stellar mass  density, with
timescales $\tau 6,  0.6$Gyrs obtained by fitting the  data at $z<1.5$
and $z>1.5$, respectively.}
\label{massdens}
\end{figure}

A more global view on the rise of the galaxy mass as a
function of cosmic time is provided by the Stellar Mass Density (SMD),
that is obtained by integrating the GSMF over all masses (we choose in
particular to integrate from log$(M_*/M_\odot)=8$ to
log$(M_*/M_\odot)=13$).  This is displayed in Fig.\ref{massdens},
where we plot as a solid line the SMD as obtained by integrating our
Schechter fiducial model.  At $z>1.3$, where we do not adequately
sample the faint end of GSMF, we rely on the extrapolation of the
slope $\alpha$ provided by our fiducial model, that corresponds to a
very mild steepening with redshift. In Tab.\ref{tabmd} we provide the
SMD values as obtained by integrating the Schechter fits as well as
those directly observed.  When compared also with other surveys,
Fig.\ref{massdens} depicts a scenario where the global stellar mass
density has evolved relatively slowly (i.e.  by a factor of two) over
the last 8 Gyrs, and more rapidly at higher redshift, with a decrease
by an order of magnitude at $z\simeq 3$.

A more direct result of our survey is the SMD in massive galaxies, defined 
as those above $M=10^{11}M_\odot$, that we directly detect in our survey 
up to $z\simeq 4$.

As we show in the lower panel of Fig.\ref{massdens}, the evolution of
the stellar mass density of massive galaxies increases fastly over the
first 3-5 Gyrs in the history of Universe, and thereafter
proceeds at a much slower pace.  It is possible to grossly reproduce
this behaviour by an exponential law, $\rho(z)\propto e^{-t/\tau}$,
characterized by two different e-folding timescales. At high redshift
($z>1.5$, i.e. look back time $>9$ Gyrs), the timescale that we derive
from our data is of the order of $0.6$ Gyrs. At later cosmic times,
the timescale is at least a factor of 10 larger: we obtain 6~Gyrs from
$z=0$ to $z=1.5$.

\begin{table}
\caption{Best fit Schechter parameters} 
\centering
\begin{tabular}{cc}
\hline 
$M^*_0$           & $11.16$       \\
$M^*_1$       & $+0.17\pm0.05$    \\ 
$M^*_2$            & $-0.07\pm0.01$ \\
$\alpha^*_0$   & $-1.18$   \\
$\alpha^*_1$  & $-0.082\pm0.033$  \\
$\phi^*_0$         & $0.0035$          \\
$\phi^*_1$      & $-2.20\pm0.18$     \\
\hline 
\end{tabular}
\label{tabSTY}
\end{table}

\begin{table*}
\caption{The Stellar Mass Density}
\begin{center}
\begin{tabular}{ccccc}
\hline
\hline
\noalign{\smallskip}
\multicolumn{1}{c}{$<z>$}  &  \multicolumn{1}{c}{Schechter} &
\multicolumn{1}{c}{Observed}  &  \multicolumn{1}{c}{Schechter} &
\multicolumn{1}{c}{Observed}\\
\multicolumn{1}{c}{ }  &  \multicolumn{1}{c}{ $8<\log{(M/M_\odot)}<13$} &
\multicolumn{1}{c}{}  &  \multicolumn{1}{c}{$\log{(M/M_\odot)}>11$} &
\multicolumn{1}{c}{$\log{(M/M_\odot)}>11$}\\
\noalign{\smallskip}
\hline
\noalign{\smallskip}
0.50 & 8.46 & $8.32^{+0.03}_{-0.03}$  &  8.10  & $7.89^{+0.13}_{-0.13}$\\
0.70 & 8.37 & $8.53^{+0.02}_{-0.02}$  &  8.01  & $8.24^{+0.07}_{-0.07}$\\
0.90 & 8.29 & $8.16^{+0.03}_{-0.03}$  &  7.93  & $7.90^{+0.10}_{-0.11}$\\
1.15 & 8.18 & $8.26^{+0.02}_{-0.02}$  &  7.83  & $7.94^{+0.06}_{-0.06}$\\
1.45 & 8.07 & $7.96^{+0.03}_{-0.03}$  &  7.70  & $7.57^{+0.09}_{-0.09}$\\ 
1.80 & 7.94 & $7.90^{+0.04}_{-0.04}$  &  7.54  & $7.50^{+0.09}_{-0.09}$\\
2.50 & 7.68 & $7.60^{+0.04}_{-0.04}$  &  7.19  & $7.16^{+0.08}_{-0.08}$\\
3.50 & 7.27 & $7.23^{+0.12}_{-0.12}$  &  6.48  & $6.60^{+0.16}_{-0.17}$\\
4.50 & ~ & $7.73                $  &  ~ & $6.44                $\\
5.50 & ~ & $7.84                $  &  ~  & $7.71                $\\
\hline 
\hline
\end{tabular}
\end{center}
\label{tabmd}
The Stellar Mass Density at different redshift bins: the second column shows
the fitted SMD (integrated from log$(M_*/M_\odot)=8$ to log$(M_*/M_\odot)=13$),
the third the observed SMD, while the fourth and the fifth columns are
the fitted and observed SMD for log$(M_*/M_\odot)>11$.
\end{table*}

\section{The comparison with theoretical $\lambda$--CDM models}

\begin{figure}
\includegraphics[width=9cm]{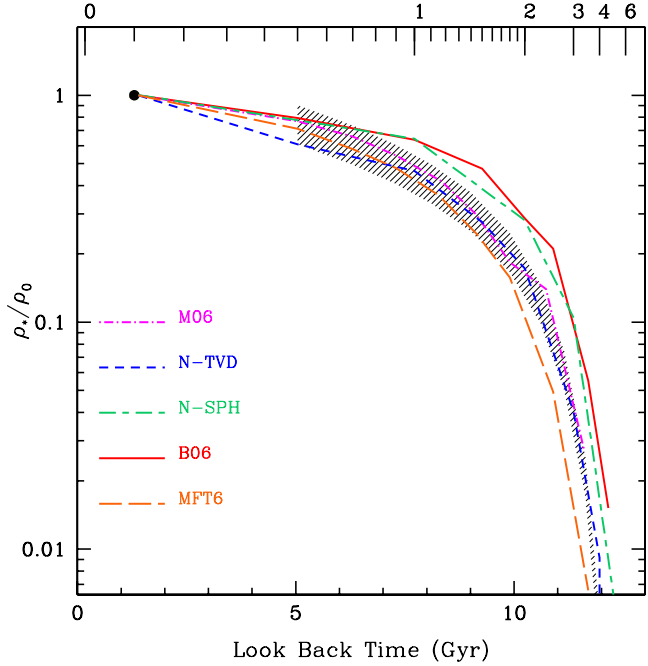}
\caption{Comparison between the observed Stellar Mass Density in
massive galaxies (defined as those above $M_*>10^{11}M_\odot$) and the
corresponding predictions by recent theoretical models, as a function
of redshift.  Both data and models have been normalized to their own
corresponding local value, for the relevant IMF. The shaded area
represent the data obtained in the present work, normalized with two
different choices of the local mass function (see text for
details). The other curves show the prediction of theoretical models,
as labelled in the legend: \cite{bower06} (B06),
\cite{menci06} (M06), \cite{monaco06} (MFT06), \cite{nagamine05a}
(N-TVD), \cite{nagamine05b} (N-SPH). }
\label{mdens_teo}
\end{figure}

In this section we compare our data with a set of recent theoretical
predictions. In particular, we have included in the comparison three
recent semianalytic models that include the feedback from AGNs on
galaxy formation, albeit with different recipes.  The first is the
latest rendition of the semianalytical Durham model (\cite{bower06},
B06 hereafter), where the feedback from AGNs is ignited by the
continuous accretion of gas on the central black hole. Such an
implementation is conceptually different from the \cite{menci06} model
(M06 in the following), where the feedback from AGN is explicitly due
to the blast wave originated during the active (luminous) phase of AGN
activity. Differences between these two models are better detailed in
\cite{menci06}.  Finally, in the semianalytic model of \cite{monaco06}
(MFT06) cooling, infall, star formation, feedback, galactic winds and
accretion onto black holes are described with a set of new simplified
models that take into account the multi-phase nature of the ISM and
its energetics.  In particular, the quenching of late cooling flows
results from the injection of energy from massive black holes that
accrete slowly (in Eddington terms) from the cooling flow itself.  A
completely different approach is provided by the hydrodynamical
simulations in a cosmological context of \cite{nagamine05a},
\cite{nagamine05b} (and references therein), which have been obtained
either with a Eulerian mesh code with Total Variation Diminishing
(\cite{ryu03}) shock capturing scheme (N-TVD in the following) or with
a SPH ``entropy formulation'' method (\cite{springel02}, N-SPH in the
following). Both models include radiative cooling and heating, uniform
UV background, supernova feedback and standard recipes for
star--formation.  In addition, the SPH simulation also include the
effects of feedback by galactic winds and a multiphase ISM, that
provides a more accurate modelling of the star--formation process.

We first compare these models with one main result of our analysis,
the evidence that the evolution of massive galaxies is relatively mild
up to $z\simeq 1.5$ and then remarkably faster. At this purpose, we
plot again in Fig.\ref{mdens_teo} the evolution of the stellar mass
density in massive galaxies as a function of redshift, normalized to
the local one, both in our data and in the models. 

Since we want to focus on the qualitative behaviour of both model and
data, and in order to minimize the impact of different IMFs, we have normalized
each model to its predicted value at $z=0$, and for the models with a
Kennicutt IMF we have used the corresponding characteristic mass $M_*$
at $z=0$.

Fig.\ref{mdens_teo} clearly shows that all models reproduce the
observed trend in the evolution of the stellar mass density, with a
slow, steady decay up to $z\simeq 1.5$ and a much faster decay
thereafter. For all models, the formation of massive galaxies is
therefore occurring at a fast pace at $z\geq 1.5-2$, and at slower
rate at lower redshift, in broad agreement with the observed
evolution. It has to be remarked, however, that the fraction of stellar mass
assembled in the high redshift phase varies by a factor of two among the 
different models, on which we shall comment later.

\begin{figure}
\includegraphics[width=9cm]{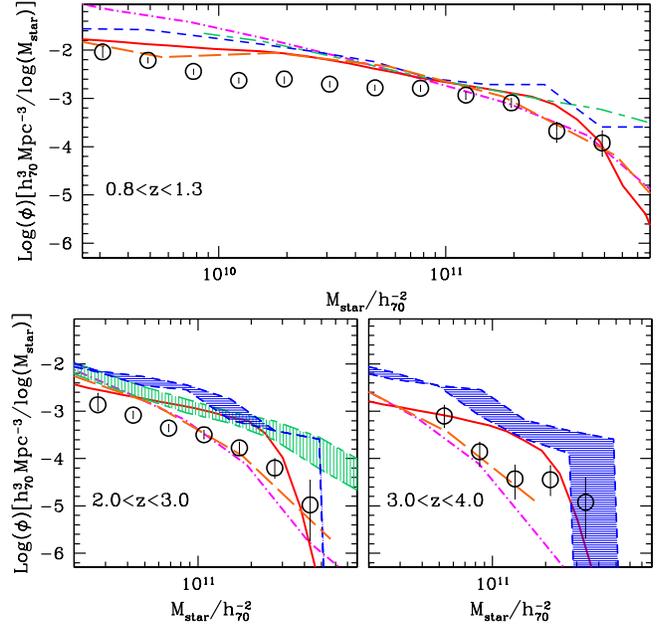}
\caption{Comparison between the observed GSMF and a set of recent
theoretical models.  As in Fig.\ref{massfun}, large circles represent
the mass--complete, Ks--selected sample. Lines refer to different models,
as in  Fig.\ref{mdens_teo}. The dashed areas represent the confidence
regions of the models by Nagamine et al 2005ab, where the limits are
the predicted mass functions calculated at $z=2$, $z=3$ and $z=4$.}
\label{MF_teo}
\end{figure}

The other important result of our analysis is that there is a
significant evolution in the density of low mass galaxies, that at
$z\simeq 1$ are about 40\% of the present--day number, and that the
GSMF is remarkably flat up to $z\simeq 1-1.3$, with a small evolution
of the slope ($\Delta \alpha = -0.1$) with respect to the local
value. A comparison with the models for this specific feature is
provided in the upper panel of Fig.\ref{MF_teo}, where we plot the
GSMF in a broad redshift bin around $z\simeq 1$. It is shown that all
the models have an unsatisfactory fit to the data, on the faint
side. Most of them fail to fit the relatively large evolution in the
density of low mass galaxies, with the result of overpredicting their
number. In addition, some of them, and in particular the M06 one,
predict a slope of the GSMF much steeper than observed.  All these
results confirm that the observed flatness of the GSMF and most of all
the significant density evolution at intermediate redshifts is a
critical feature, difficult to reproduce by current theoretical
models.

We also notice that, always at $z\simeq 1$, most of the models roughly
predict a correct number of massive galaxies, with the exception of
the two hydrodynamical simulations. The consequence of this comparison
is that the ratio between massive and low mass galaxies (whose
evolution provides the ``downsizing'' scenario) is not reproduced by
any model: it is interesting to remark that such a failure is due to
the overprediction of low mass galaxies, and not to an underprediction
of massive galaxies.

In the bottom panels of Fig.\ref{MF_teo} we also compare the
theoretical GSMFs to the observed ones in the highest redshift
bins. The comparison in the redshift range $2<z<3$ is probably the
most statistically meaningful, albeit anyway limited to the higher
mass regime, and we briefly concentrate on it.  It is shown that,
at variance with the overall agreement found at $z\simeq 1$, the
considered model span a wide range of predicted densities. Typically,
hydrodynamical models tend to overpredict the observed data, while the
semianalytical renditions appear to be closer to the data. In the detail,
those that include a ``QSO-like'' feedback from
AGNs (M06 and MFT06) tend to underpredict the data, and those
with a ``radio-like'' AGN
feedback (B06) tend to overpredict the data.  While the disagreement is
apparently large (by nearly an order of magnitude) as far as the
number densities are concerned, it must be remarked that the effect is
emphasized by the steep slope of the exponential tail of GSMF, since
the offset in mass between the theoretical predictions and the data is
within a factor of two. Since hydrodynamical models do not include
feedback from AGNs, while ``QSO-like'' feedback from AGNs is
particularly efficient in massive halos at high redshift, it is
tempting to ascribe the difference to this specific feature, although
 many details of the galaxy formation models, especially the 
cooling and infall of gas in the infall-dominated halos at high 
redshift, may play a  fundamental role as well.

\section{Massive galaxies at $z>4$}

As we have described in the previous sections, our statistical
analysis is limited to $z<4$ to take into full account the various
effects of incompleteness in the sample. However, as clearly shown in
Fig.\ref{massz}, our sample includes several galaxies at $z>4$. Some
of these are ``drop--out galaxies'', that are brighter than $m_z=26$
and detected in $Ks$, the two requirements that we adopted in plotting
Fig.\ref{massz}.  As such, our sample does not include many other ``I
drop--out galaxies'', likely at $z>5$, that are detected in the GOODS
images, since they are either fainter than $m_z=26$ or undetected in
$Ks$. These objects do not make up a complete sample
and are therefore excluded from our current analysis.

Most notably, however, we have a few very massive candidate galaxies
at $z>4$ in our sample. It is interesting and worthwhile to carry on a
detailed analysis of these objects, especially because they give a
significant contribution to the mass density of galaxies.  In fact,
when we estimate the stellar mass density for the galaxies with $z>4$
(Fig.\ref{massdens}), we note that it is very high, despite it is a
lower limit: this is more evident in the lower panel of the
figure where only the mass density of the very massive galaxies is
shown. Few very massive galaxies are responsible for such high value
of the mass density: at $z=3-4$ two massive galaxies give $34\%$ of
the total mass density, while at $z=5-6$ the mass density of only
three objects is $94 \% $ of the total value.

\begin{figure}
\includegraphics[width=9cm]{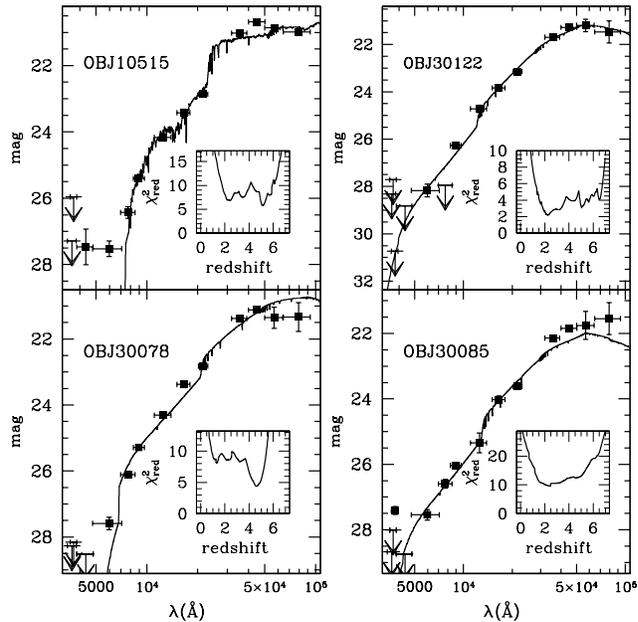}
\caption{Left: Spectral Energy Distributions of two galaxies with
photometric redshift $>4$, that would result very massive $M_*\simeq
5\times 10^{11}M_\odot$. In the inset, the $\chi^2$ as a function of
redshift of the photometric redshift is shown. Right panel: Spectral
Energy Distributions of two objects with similar colours and
uncertainty in redshift, that are assigned a photometric redshift
$\simeq 2$.}
\label{z4}
\end{figure}

It is interesting to investigate what kind of objects are these.  If
we select the galaxies with log$(M_*/M_\odot)>11$, we find three
objects, as shown in Fig. \ref{massz}; two of them are Ks and
$Z_{850}$-selected, and one is only Ks-selected.  These objects are typically
quite red for their redshifts, with a rest--frame $U-V \sim 1.5$. The
best-fitting templates for these objects are either passively evolving
models, characterized by a very short timescale for star formation
$\tau$ with $E(B-V)\sim 0$ or by a constant star formation model with
large amount of dust, $E(B-V) > 0.5$.  Two of these objects are
displayed in the left panel of Fig. \ref{z4}.  It is important to
stress that the redshift determinations of all these objects are very
uncertain, as it is shown by the $\chi^2$ as a function of redshift in
the same figure. In practice, on the basis of the available data one
can only conclude that these objects are most likely at $z>2$, but a
sound redshift determination is not available. In particular, other
redshift solutions can be find with a similar probability at $z=2-3$.

We note that these objects might resemble the object presented in
Mobasher et al. 2005, that is also a candidate massive, passively
evolving galaxy at $z\simeq 6.5$.  We have also extracted from our raw
catalog the photometry of the Mobasher et al. (2005) object (that is
slightly fainter than our limiting magnitude in $Ks$), to check how it
is classified with our photometry, that is based on a different image
set (Mobasher et al. (2005) used a combination of UDF images taken with
ACS and NICMOS). With our photometry, such an object is quite similar
to our $z\simeq 2$ candidates in Fig. \ref{z4}, with a flat probability
distribution for the photometric redshift from $z=2$ to $7$ and the
best fit solution at $z=2.25$ with $E_{B-V}=1.1$: in this case the
stellar mass turns out to be $2\cdot 10^{11}M_{\odot}$, quite typical
at this redshift. 

At this point, it is interesting to seek for similar galaxies in the
GOODS--MUSIC sample, i.e. galaxies that have a comparable spectral
distribution, best-fitting template around $z=2-3$ and with a
probability distribution extending up to $z\simeq 5-6$.  Such objects
indeed exist in our sample: we find in total three objects, a number
comparable to the number of the massive object at higher redshift. Two
of them are shown in the right panel of Fig. \ref{z4}.  The fraction
of objects with uncertain redshifts is negligible at $z=2-3$ where we
have estimated MF and mass density, such that we do not expect that
they introduce a significative uncertainty in the statistical
analysis. In any case, the effect of such uncertainty is taken into
account by the Monte Carlo estimate of the error budget in the GSMF.

We can draw two different conclusions from this exercise.  First,
although we cannot exclude that some of these objects are actually
very massive galaxies at $z>4$, we have to await for a more robust
determination of their redshift before including them in a firm
estimate of the stellar mass density at high redshift.  

Second, the properties of the GSMF can give some hints on the nature
of these objects.
Probably the most important result of our analysis is that we can
exclude that these objects are typical in a statistical sense. Our
sample is indeed complete for passively evolving galaxies of stellar
mass $M_*>10^{11}M_\odot$ up to $z\simeq 4$, and we can firmly
estimate a decrease of at least a factor of ten in their density at
such high redshift. A scenario where most of present--day early
galaxies formed at even higher redshift $z>4$ and evolved passively
thereafter is therefore excluded by our analysis, such that we can
conclude that the large mass density contained in passively evolving
systems that would arise from our $z>4$ candidates is either a
statistical fluctuation, or results from incorrectly assigned
photometric redshifts. Future analysis will hopefully clarify this
point.


\section{Summary and Discussion}

In this work we have presented an analysis of the evolution of the Galaxy
Stellar Mass Function (GSMF) and of the corresponding stellar mass
density up to $z\simeq 4$.  It has been obtained from the GOODS-MUSIC
sample, a $Ks$--selected catalog of 2931 galaxies with 14 bands
photometry, extending from $0.35$ to $8\mu$m, extracted from the public data
of the GOODS-South survey.  We have derived accurate stellar masses
for this large sample of galaxies, adopting the standard technique of
fitting spectral synthesis models (Bruzual \& Charlot 2003) on the
multiwavelength data. To compare with previous works, we have used the
Salpeter IMF, ranging over a set of metallicities (from $Z=0.02
Z_\odot$ to $Z=2.5 Z_\odot$ and dust extinction ($0<E(B-V)<1.1$, with
a Calzetti extinction curve).  On this data set, we have computed the
Galaxy Stellar Mass Function and the resulting stellar mass density
with the usual $1/V_{max}$ formalism. To provide a visualization of
the GSMF free of the well known biases of the $1/V_{max}$ approach
(namely, the sensitivity to large scale structures and the arbitrarity
of the binning), we also provide a global fit to the data with a
Schechter function with smooth, redshift--dependent free parameters
(``STY approach'').

We have carefully discussed the selection effects that play a role in
our $Ks$-selected sample, and that must be taken into account in the
analysis.  We have shown that, while we can detect
massive ($M_*>10^{11}M_\odot$) passively evolving galaxies up to
$z\simeq 4$, our sample becomes progressively biased against
star--forming, dust-enshrouded objects with $E(B-V)>0.5$ already at
$z>2$. Since there are evidences that these objects do exist at $z<2$,
it is possible that our census of galaxies at $z>2$ is incomplete, and
that our estimates in this redshift range (as in any $Ks$--selected
sample) should be considered as a lower limit.

The major results that we have found with these data are the following:

- We show that the inclusion of the Spitzer data (from 3.5 to 8
$\mu$m) significantly improves the reliability of the mass estimate at
$z>2$, as expected. At lower redshift, there is no significant
improvement and the observed scatter is consistent with that induced
by the model degeneracy.

- We confirm the well known trend of global decline of the stellar
mass density with redshift. The total mass density is about a factor
2$\times$ lower than local at $z\simeq 1$, and about 10\% of the local
at $z\simeq 3$.

- We compute the GSMF in several redshift bins, from $z=0.4$ to $z=4$,
and we show that a simple scaling of the Schechter parameters is able
to provide a smoothly evolving rendition of the GSMF.

- Restricting only to galaxies with $M_*>10^{11}M_\odot$,
their mass density evolves relatively mildly up to $z\simeq 1-1.5$. At
$z\simeq 1$, their integrated mass density is about 50\% of the local
value.  At $z>1.5$, massive galaxies become to evolve much faster,
such that at $z\simeq 3$ they provide at most one tenth of the local
density. This trend may be described by two exponential e--folding
times, i.e. $\rho(z)\propto e^{-t/\tau}$, with $\tau\simeq 6$Gyrs at
$z<1.5$ and $\tau\simeq 0.6$Gyrs at $z>1.5$, although this number may
be biased low because of the selection effects described above.

- As far as low mass galaxies are concerned, we show that the GSMF
remains remarkably flat (it steepens by less than 0.1 for each unit
redshift) up to $z\simeq 1-1.3$, the highest redshift bin where it can
be reliably measured. At the same time, a sensible decrease occurs in
the number of low mass galaxies: the density of galaxies with $M\simeq
10^{10}M_\odot$ at $z\simeq 1$ is 4 times lower than the local one.

- In this context, we finally show that there is a clear evidence for
a differential evolution of the Galaxy Stellar Mass Function up to
$z\simeq 1.5$, with less massive galaxies evolving more than massive
ones. Such a trend is evident in Fig.\ref{massfun}, that shows that
the GSMF in the redshift bins up to $z\simeq 1.3$ remains
progressively closer to the local one for increasing stellar masses,
and it is also substantiated by two quantitative evidences. First,
from $z=0$ to $z\simeq 1.5$ the total stellar mass density decreases
more than that due to massive galaxies only (i.e. those with
$M^*>10^{11}M_\odot$). More intriguingly, the STY fit to the overall
evolution of the GSMF shows that the characteristic mass $M^*$
increases up to $z\simeq 1.3$, reflecting the change in the shape of
the GSMF, that from $z\simeq 0$ to $z\simeq 1.3$ is progressively more
skewed toward higher masses.

- After $z\simeq 1.5$, however, the evolution of the mass density
contained in massive galaxies is much faster, as shown by
Fig.\ref{massdens}, showing that this is their main epoch of
formation.

It is straightforward to see that such differential evolution of the
GSMF is a natural consequence of the ``downsizing'' scenario for
galaxy evolution, that has been found in many independent
surveys. Following the original definition by Cowie et al. 1996, it
has been shown by several authors (e.g. \cite{brinchmann00},
\cite{fontana03}, \cite{perez05}, \cite{feulner05}) that the specific
star--formation rate increases with redshift, showing that massive
galaxies become progressively more actively star--forming as redshift
increases. By looking at the GSMF, we observe the consequences of such
a trend: more massive systems form in a vigorous phase at high
redshift, that is largely complete at $z\simeq 1-1.5$, such that the
corresponding section of the GSMF is already close to the local
value. Lower mass systems, on the contrary, continue to grow their
stellar content at even lower redshifts, such that the increase of the
faint side of the GSMF is large also at low $z$.

We remark that this picture is not plagued by the biases against the
detection of dusty massive galaxies at high $z$ that exist in our
sample, as in any $Ks$ complete one. As we discussed in the text, we
might be missing high redshift, star--forming dusty galaxies with
large extinction, but we are essentially complete with respect to
passively evolving, dust-free galaxies of $M_*>10^{11}M_\odot$ up to
$z\simeq 4$.  The low density of high redshift massive systems
provides decisive evidence that the formation and assembly of local,
massive bulges and ellipticals did not form in a single phase at very
high redshift.

We have explored whether such a picture is qualitatively consistent
with the predictions of most recent theoretical models for galaxy
formation. We have compared our data with a set of theoretical models,
including semianalytic models (\cite{bower06}, \cite{menci06},
\cite{monaco06}) as well as hydrodynamical ones (\cite{nagamine05a},
\cite{nagamine05b}). All these models predict a relatively mild
evolution of the stellar mass density contained in massive galaxies
from $z=0$ to $z\simeq 1.5$, that is broadly consistent with the
observed data, and a much faster evolution at higher $z$, again in
agreement with the data.

For all the models, however, at intermediate redshifts the match with
the slope and the normalization of the GSMF at intermediate or low
masses ($M_*\simeq 10^{10} M_\odot$) is still critical, when not even
poor. Interestingly, this implies that the ``downsizing'' scenario
(that is based on the evolution ratio between massive and low mass
galaxies) is not reproduced by any model because of the overprediction
of low mass galaxies, and not because  of an underprediction of the
massive ones.

At high redshift, the detailed agreement with the observed data is
still sensitive to the description of the physical processes inserted
in the models. Hydrodynamical models tend to overpredict the observed
mass density, as already noted by the authors (\cite{nagamine04}),
while the semianalytic models that include feedback from AGNs are
closer to the data.

In conclusion, the new data presented in this work provide an overall
description of the rise of the stellar mass, and in particular of that
residing in massive galaxies, in which about half of such stellar mass
appear to have been assembled during the first 2-4 Gyrs after
recombination, followed by a milder increase over the remaining cosmic
time. Although encouraging, the comparison with the theoretical
expectations provides evidences that some fundamental physical
processes, likely affecting both low and high mass galaxies,
are still incorrectly represented, and that the density of massive
galaxies at high redshift is indeed a very useful test for these
models. 

At this purpose, more effort is needed to improve the reliability
of the observational estimates at intermediate and high redshift. From
the observational point of view, larger spectrophotometric surveys on
independent fields are definitely necessary to smear out the effects
of sample variance. These might in particular affect two of our major
results, namely the increase of the characteristic mass $M^*$ up to
$z\simeq 1$ and the amount of massive galaxies at $z>2$.  In addition
to such (obvious) caveats, several systematic effects are still to
be properly minimized. To mention a few, those related to the choice
of the IMF and to the differences that may arise from the treatment of
post--AGB stars in spectral synthesis models (\cite{maraston05}), and
the possible contribution of dust--enshrouded galaxies to the overall
mass budget.

In addition, these findings rise important questions about the
physical processes that led to the rise of the stellar mass density in
massive galaxies: what is the physical nature of the galaxies that
contribute to the stellar mass density at high redshifts, and what is
the physical mechanism that drove this rise, i.e. how much it is
related to star--formation occurring within the observed galaxies as
compared to the contribution from merging processes.  Although
present--day surveys are starting to explore these issues, these will
remain among the more challenging questions of the next years.

\begin{acknowledgements}
We thank the referee, Niv Drory, for his useful suggestion, that led
to a significant improvement of the paper. We thank K. Nagamine and
R.C. Bower for having allowed us to use their latest models.  We also
acknowledge fruitful discussion with A. Cimatti, M. Dickinson,
L. Moustakas, L. Pozzetti, A. Renzini and G. Zamorani.  It's a
pleasure to thank the whole GOODS Team for providing all the imaging
material available worldwide. Observations have been carried out using
the Very Large Telescope at the ESO Paranal Observatory under Program
IDs LP168.A-0485 and ID 170.A-0788 and the ESO Science Archive under
Program IDs 64.O-0643, 66.A-0572, 68.A-0544, 164.O-0561, 163.N-0210
and 60.A-9120.
\end{acknowledgements}

\end{document}